\documentclass[global,twocolumn]{svjour}
\usepackage{graphics}
\usepackage{color}
\newcommand{\LN}{LN$_{\textrm{2}}$}
\usepackage[cdot,squaren, thinspace,thinqspace]{SIunits}
\usepackage[textsize=scriptsize]{todonotes}
\journalname{myjournal}
\begin{document}
\title{Liquid-nitrogen cooled, free-running single-photon sensitive detector at telecommunication wavelengths}
\author{M. Covi\inst{1} \and B. Pressl\inst{1}\and T. G\"unthner\inst{1} \and 
 K. Laiho\inst{1}\thanks{\emph{e-mail:} kaisa.laiho@uibk.ac.at},
 S. Krapick \inst{3}, C. Silberhorn \inst{3} \and G. Weihs\inst{1,2} 
}                     
%
%
\institute{Institut f\"ur Experimentalphysik, Universit\"at Innsbruck, Technikerstra\ss e 25, 6020 Innsbruck, Austria \and Institute for Quantum Computing, University of Waterloo, 200 University Ave W, Waterloo, ON, N2L 3G1, Canada 
\and Applied Physics, University of Paderborn, Warburger Stra\ss e 100, 33098
Paderborn, Germany }
\date{\today}
%
\maketitle
\begin{abstract}
The measurement of light characteristics at the single- and few photon level plays a key role in many quantum optics applications. Often photodetection is preceded with the transmission of quantum light over long distances in optical fibers with their low loss window near \unit{1550}{\nano\meter}. 
Nonetheless, the detection of the photonic states  at telecommunication wavelengths via av\-a\-lanche photodetectors has long been facing severe restrictions. Only recently, demonstrations of the first free-running detector techniques in the telecommunication band have lifted the demand of synchronizing the signal with the detector. Moreover, moderate cooling is required to gain single-photon sensitivity with these detectors. Here we implement a liquid-nitrogen cooled neg\-a\-tive-feedback avalanche diode (NFAD) at telecommunication wavelengths and investigate the properties of this highly flexible, free-running single-photon sensitive detector. Our realization of cooling provides a large range of stable operating temperatures and has advantages over the relatively bulky commercial refrigerators that have been used before. We determine the region of NFAD working parameters most suitable for single-photon sensitive detection enabling a direct plug-in of our detector to a true photon counting task.
\end{abstract}

\section{Introduction}
\label{sec:1}
The observation of a light quantum is a demanding task since the detector should be able to resolve this tiny amount of energy  via the heat absorption or the photo-electric effect and in a noiseless fashion convert it to a measurable electric signal. In addition to this, quantum features are extremely fragile and easily lost by any experimental imperfections prior to and at the detection. Nonetheless, in the past years, several photo-detector technologies have demonstrated the sensitivity to measure at the single-photon level \cite{Eisaman2011}. Moreover, many quantum optics applications require photon counters with at least partial photon-number resolution \cite{Kardynal2008}. In such experiments, superconducting detectors \cite{Lita2008,Gerrits2010} as well as visible light photon counters \cite{Waks2004} can be exploited with high efficiencies---however, at the cost of having to cool to temperatures of a few Kelvin. At room temperature only a few options are available, for example photo-multiplier tubes \cite{Hong1986} that suffer from large losses or hybrid detectors \cite{Bondani2009}  that require specific reconstruction techniques.

For photon counting tasks the most widely used detectors are avalanche photo diodes or shortly  \emph{click detectors} that are in Geiger mode sensitive to single photons but unable to resolve various photon numbers \cite{Silberhorn2007}. This is due to the fact that a whole bunch of carriers is released after a photon impinges on the detector in order to gain a detectable electrical signal \cite{Buller2010}. This technology for the detection of light at the single-photon level is well-established and commercially available. In the region of visible light, click detectors can reach high quantum efficiencies and measure with a very low number of spurious counts \cite{Hadfield2009}. Additionally, in connection to a fiber-optic beam splitter network these detectors have lately been utilized as photon-number resolving detectors via time multiplexing \cite{Achilles2003,Fitch2003}.

Regarding a particular quantum optics application the selection of a suitable detector starts by determining the wavelength range of the investigated radiation. In the range of telecommunication wavelengths, which are ideal for long-distance transmission of light in optical fibers, click detectors have suffered from several drawbacks. Until recently these detectors have required narrow time\-gating  to suppress spurious counts, in other words, these detectors are active only for a short duration of time and the measured signal has to be synchronized with the gate. In a free-running mode -- first demonstrated for the telecommunication band in Refs.~\cite{Thew2007,Warburton2009a,Warburton2009} -- this synchronization can be omitted \cite{Yan2012,Korzh2014}. The better temporal availability directly offers the possibility to measure at higher rates. Additionally, free-running detectors make time-multiplexing and photon-counting via a fiber network more practicable at telecommunication wavelengths.

Apart from the wavelength sensitivity and efficiency, the temporal characteristics of a detector such as dead time, timing jitter, as well as the amount of spurious counts in terms of the dark counts and afterpulses are the figures of merit of click detectors \cite{Cova1996}. While wavelength region and quantum efficiency mostly depend on the semiconductor's properties the above characteristics are also influenced  by the detector realization such as the stochastic process describing the carrier avalanche. However, also the external electronics can have great impact especially on the temporal characteristics of the detector \cite{Warburton2009}. For example the technique used for quenching the avalanche often limits the dead time of the detector and the read-out electronics the timing jitter. The rate of spurious counts often strongly depends on the bias voltage \cite{Yan2012}, which also sets the internal (avalanche) amplification. For a high signal-to-noise ratio we would like a high efficiency and low spurious counts, or in other words the lowest possible noise equivalent power (NEP, or better noise-equivalent photon count rate). The NEP has to be controlled, otherwise the detector looses its single-photon sensitivity  \cite{Warburton2009}. To complicate things, these properties are temperature dependent and can thus be manipulated by cooling the detector. Cooling typically increases the detector efficiency at a given dark count rate \cite{Yan2012}. However, care has to be taken since this comes at the cost of an increasing probability of afterpulses \cite{Korzh2014} and a shift in the wavelength sensitivity of the detector towards shorter wavelengths.

Here, we present  a highly  practical realization of a liquid-nitrogen (\LN) cooled single-photon sensitive detector at telecommunication wavelengths. It is based on a commercially available fiber-coupled neg\-a\-tive-feedback avalanche diode (NFAD) (PNA-300-1) from Princeton Lightwave. The integrated feedback resistor makes this diode intrinsically passively self-quenched with a well-de\-fined pulse form and a consistent reactivation time \cite{Lunghi2012}. Our implementation of deriving an output signal follows Yan et al.~\cite{Yan2012}. Our solution of cooling the detector with liquid nitrogen has several advantages. A wide range of temperatures can be chosen, the system is compact and easily movable. It does not require any bulky cooling apparatus and it efficiently prohibits humidity from condensing on the detector heads with minimal preventions taken. In contrast to other methods that may necessitate combination of coolers  such as thermoelectric cooling and water cooling, \LN  \ requires only a single step and can be implemented at low cost in laboratories, where \LN  \  is already available.

This paper is organized as follows. In Sec.~\ref{sec:2} we present the chosen cooling technique and study its temperature characteristics.
In Sec.~\ref{sec:3} we investigate the optical properties of our implementation. We examine both the efficiency of the detector as well as the NEP at different temperatures in terms of the dark count rate and show how to suppress the effects of afterpulsing. After finding the working parameters for the most suitable operating point we apply our detector in a true quantum optics task for coincidence counting. Sec.~\ref{sec:4} summarizes our results and highlights the findings of our studies.

\section{Detector realisation}
\label{sec:2}

\begin{figure}
    \centering
     \includegraphics[width=0.8\columnwidth]{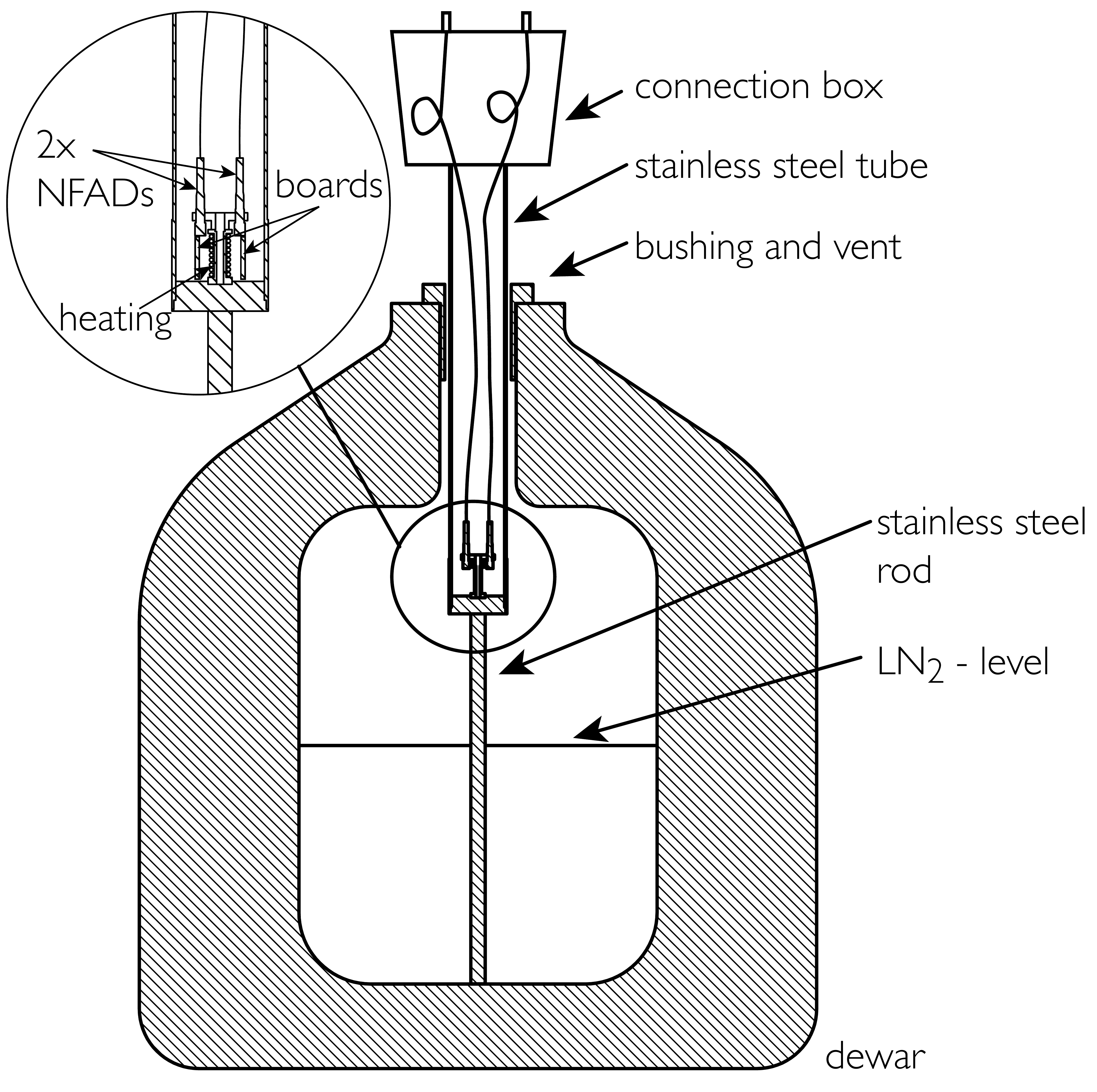}
    \caption{Cross-section of the detector cooling setup. For more details see the main text.}
    \label{fig:2.1}     
\end{figure}

Our realization of a free-running, single-photon sensitive detector is based on \LN \ cooling. A schematic picture of our implementation having a height of approximately \unit{80}{\centi\meter} is shown in Fig.~\ref{fig:2.1}. We put our fiber-coupled NFADs in a capsule placed on top of a \unit{34}{\centi\meter} long metal rod of stainless steel, which is set into a partially filled dewar of \LN \ such that the  detectors sit above the surface of the liquid. We can pack two NFADs with small circuit boards in one capsule including a heating coil placed in between them as depicted by the inlay in Fig.~\ref{fig:2.1}. All  wiring is routed via a stainless steel tube to a connection box at the top of the tube. A  vented bushing allows evaporated nitrogen to escape the dewar and holds the construction straight.

\begin{figure*}[!htb]
    \centering \includegraphics[width=0.8\textwidth]{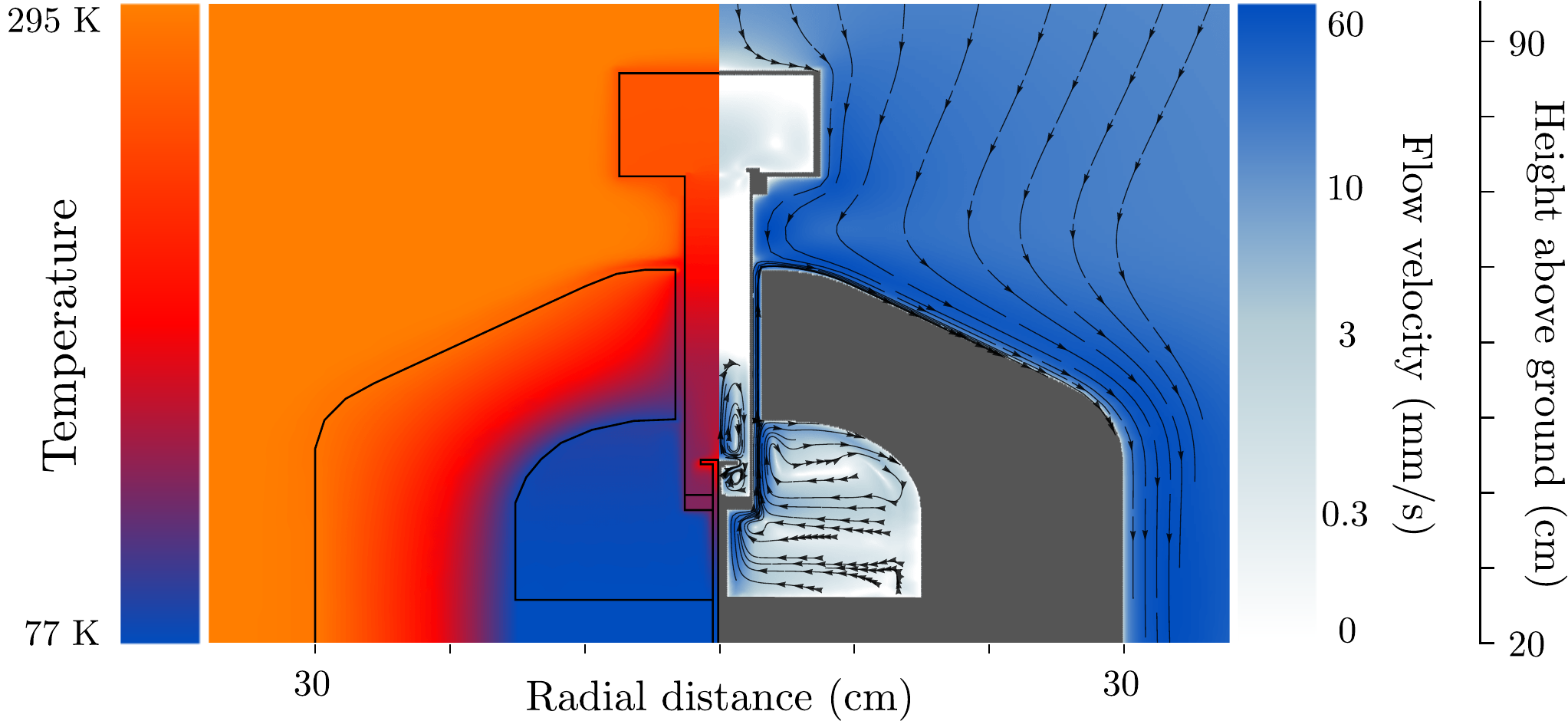}
    \caption{Simulation of temperature distribution (left) and the heat transfer (right) in the \LN \ filled dewar and its surroundings, when heating is applied. The strength of the stream is shown by the flow velocity. The pathlines illustrate the traces of individual fluid particles. The time difference between two arrowmarkers within one trace is kept constant.}
    \label{fig:2.2}     
\end{figure*}

We employ simulations with a commercial solver to investigate the heat transport inside the dewar and at its opening. This is especially important as the NFADs have to be protected from condensed water. In order to gain the proper parameters for the simulation, we make a one day test run, in which the temperature of the detector housing is kept constant at \unit{-80}{ \celsius}. We observe an \LN \ evaporation rate of approximately \unit{150}{\gram \per \hour}.  At this rate a single dewar fill of \unit{15-20}{\kilo\gram} yields an operating time of \unit{96}{\hour}. A small amount of condensed water, which forms during the test run near the vent, can be easily collected with a tissue placed loosely at the neck of the dewar. The long term test also shows that a heating power of approximately \unit{6.5}{\watt} is required to keep the operating temperature stable.

The simulated dewar shape and size correspond to the dimensions of our real dewar with a \unit{35}{\litre} interior space filled to a specific level with \LN{}. The boundary conditions are the measured evaporation rate, an environment temperature of \unit{25}{\celsius} and \unit{1}{\bbar} pressure. To simulate the effect of the heating coil, \unit{7}{\watt} of power is deposited in the heating area inside the detector capsule. 
 
Our simulation in Fig.~\ref{fig:2.2} firstly shows as expected that with the chosen parameters a temperature of \unit{199}{K} can be achieved on the horizontal plate in the detector capsule, where the NFAD-detectors are attached. Secondly, it can be clearly seen that the heat is transported not only via conduction but also via convection that causes a small eddy inside the dewar. More interestingly, there is a continuous flow of evaporated nitrogen from the dewar to the outside, which prohibits the condensation of water and the forming of ice on the detector house and the cooling rod. A similar eddy inside the detector capsule cools and protects the NFADs itself.

Although the strength of the heat transport is strongly dependent on the amount of \LN, a stable operating temperature can be achieved by a commercial temperature controller that regulates the heating power. In a second \unit{3}{\hour} long test at \unit{-80}{\celsius} we measure the temperature stability of our diodes and find that the operating temperature can be set with a stability of \unit{\pm0.002}{\kelvin}.

\section{Experimental investigations}
\label{sec:3}

In order to determine the most suitable working range of our single-photon sensitive detectors we investigate their main figures of merits.  Although each individual detector has its own specific characteristics we present here the figures of merit for one single NFAD. 

\begin{figure}
    \hspace{-6mm}
     \includegraphics[width=1.23\columnwidth]{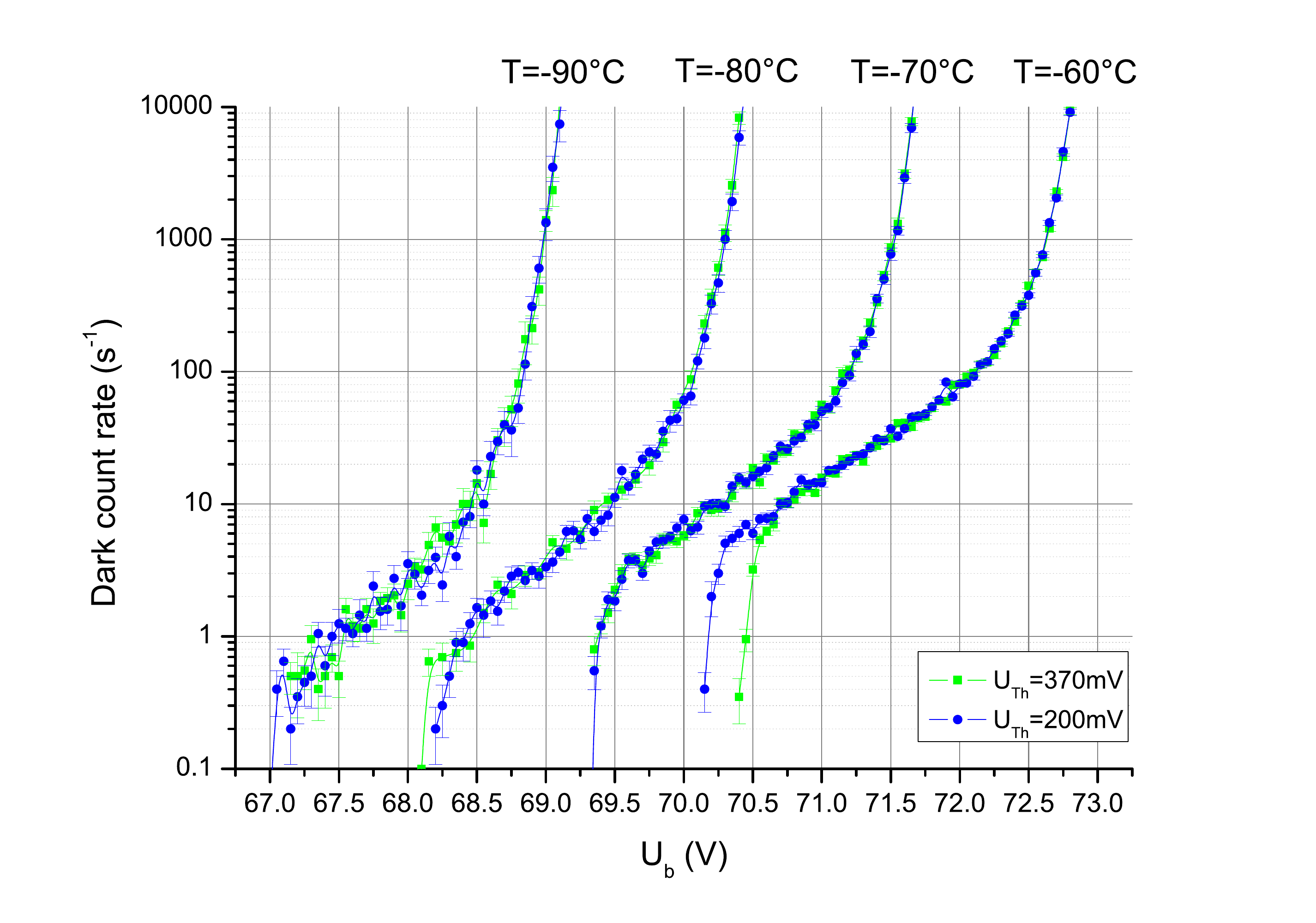}
    \caption{Dark counts in terms of the bias voltage  $U_{\textrm{B}}$ for a low and high threshold voltage (shortly $U_{\textrm{Th}}$) at different operating temperatures (marked with $T$) from \unit{-60}{\celsius} to \unit{-90}{\celsius}.  }
    \label{fig:3.1}    
\end{figure}

We start by investigating the dark count rate with respect to the bias voltage and the detector's temperature. The electrical response of the NFAD is discriminated at a desired threshold level and measured via a multi-channel time-to-digital converter (TDC) that serves as a counter of the detected clicks. As the electrical response of the NFADs shows a highly complicated pulse shape, we choose to measure the dark counts with two different threshold values for the TDC counting the clicks. Our results in Fig.~\ref{fig:3.1} show the working range of the detector in terms of the bias voltage and the achieved dark counts at various temperatures. We further note that the threshold value can be chosen rather freely between two values we used in our measurements as they both deliver the same tendency of the measured counts. 

\begin{figure}[!h]
\begin{picture}(167,167)%
   \put(0,0){ \includegraphics[width=1.0\columnwidth]{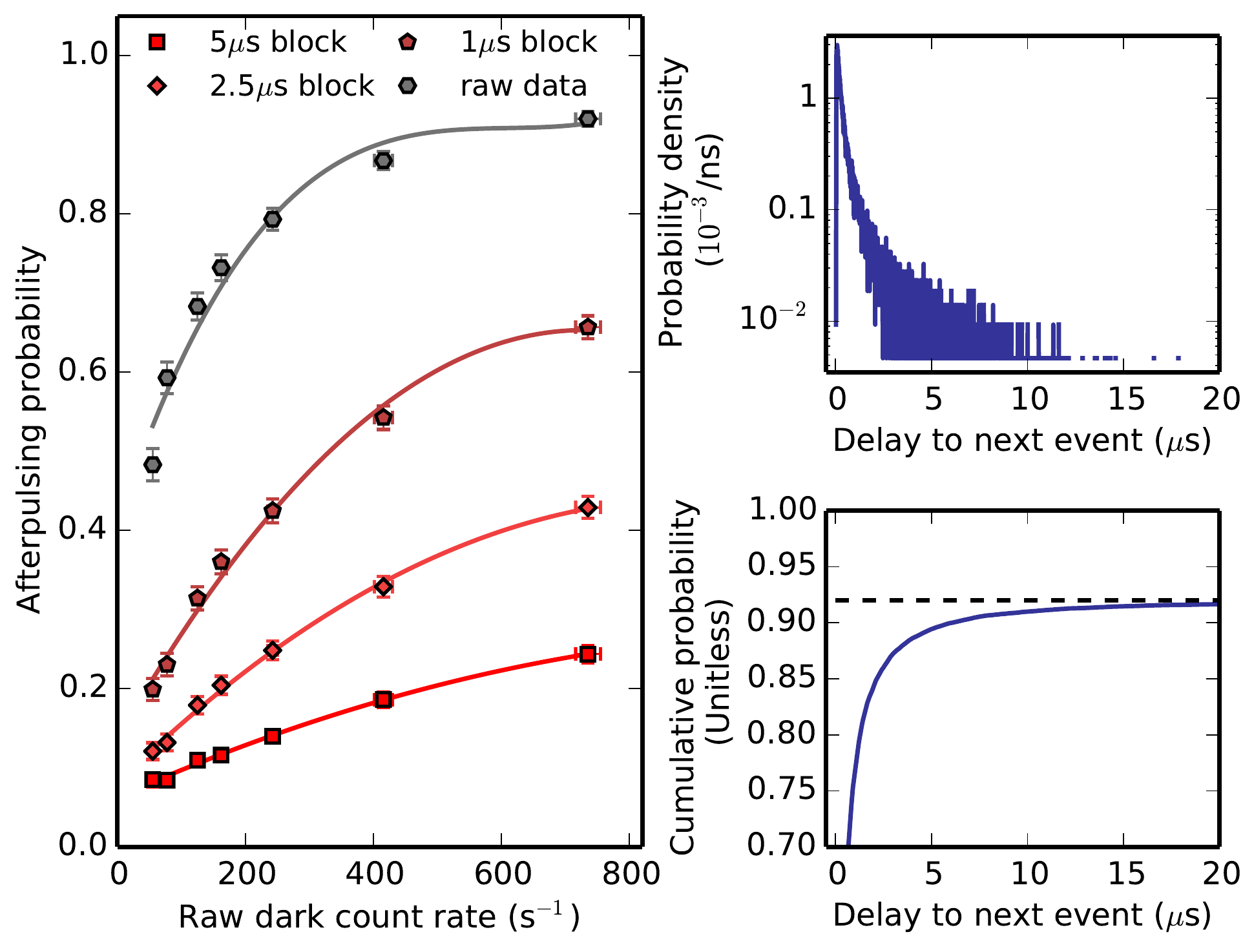} }
   \put(32,152){\scriptsize(a)}
      \put(200,152){\scriptsize(b)}
            \put(200,40){\scriptsize(c)}
   \end{picture}
 \caption{Afterpulsing characteristics of the detector at  \unit{-60}{\celsius}. (a) The raw data shows a large afterpulsing probability. By postprocessing the data with a proper blocking time the afterpulsing probability can be largely suppressed. While measured values are illustrated with symbols, the solid lines present a guide for the eye. (b) Temporal distribution of the registered events at the highest measured raw dark count rate in (a) shown up to \unit{20}{\micro\second}. (c) The cumulative probability of the temporal distribution in (b) shows that a blocking time of about \unit{5}{\micro\second} is justified to suppress the afterpulsing at this temperature. The dashed line marks the extracted raw afterpulsing probability.}
    \label{fig:3.2}      
\end{figure}
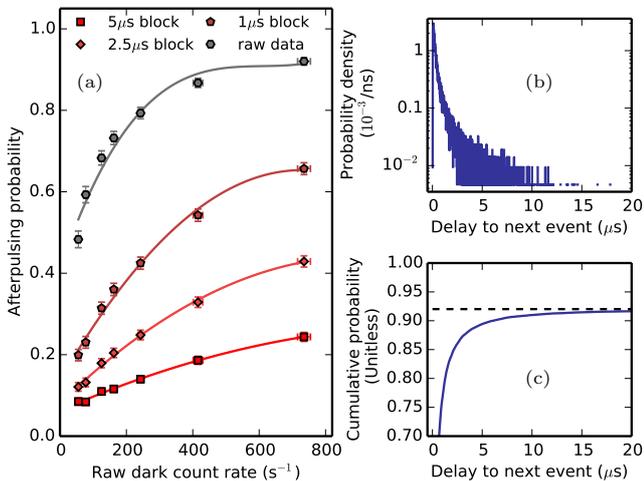

Apart from controlling the dark counts also afterpulsing that introduces artificial detection events has to be governed. For this purpose, we measure the dark counts in a time-resolved manner with \unit{200}{\milli\volt} threshold, which is kept at this level in all the coming measurements, and investigate their temporal characteristics. We label all clicks happening in proximity of another event as afterpulses and record afterpulses until a time given by 10\% of the inverse of the raw dark count rate is passed.
We estimate the afterpulsing probability as a ratio of those events to all registered detection events. Our results in Fig.~\ref{fig:3.2}(a) show that the detector is highly influenced by afterpulsing already at  \unit{-60}{\celsius} and would be practical only at very low dark count rates or bias voltages. In Fig.~\ref{fig:3.2}(b) we demonstrate a typical temporal distribution of the detection events proceeding a registered click, which indicates that there is an increased probability of detecting another event just after some tens of nanoseconds from a previous one. Fig.~\ref{fig:3.2}(b) also indicates that to counter the afterpulsing, a blocking time should be applied, which drastically diminishes the afterpulsing effect, as seen in Fig.~\ref{fig:3.2}(a). The cumulative probability in Fig.~\ref{fig:3.2}(c) of the temporal distribution [Fig.~\ref{fig:3.2}(b)] can be used  to quantify the required blocking time. 

Applying blocking means that we let the detector run freely and post-select only those events, before which no event was detected within the selected time frame. Of course, the blocking time  limits the useful signal rate to a value given by its inverse.
Our investigations of afterpulsing at lower temperatures show a very similar behavior to that recorded at  \unit{-60}{\degree}. However, due to increasing afterpulsing longer blocking times are required to suppress the afterpulsing probability to the same level. This way the detector can still be used down to temperatures of \unit{-90}{\celsius}. The appropriate blocking time depends on how large  an afterpulsing probability can be accepted. The trade-off between the highest acceptable afterpulsing probability and the shortest blocking time required delivers a clear upper bound for the raw dark count rate at a chosen temperature, below which the detector has to be operated.

\begin{figure}[!t]
\hspace{-5mm}
\begin{picture}(165,120)%
   \put(0,0){
    \includegraphics[width=1.05\columnwidth]{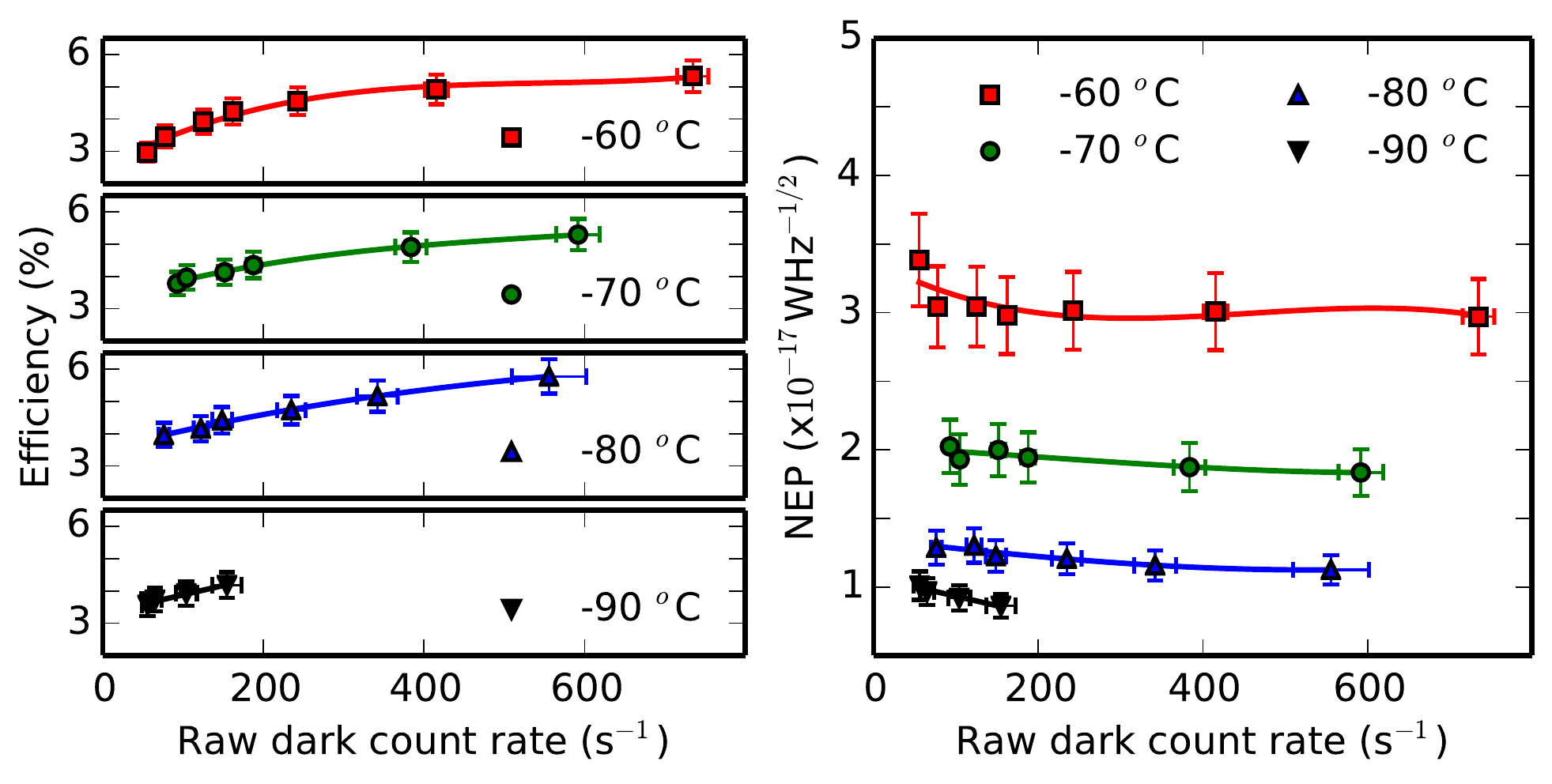}}
             \put(22,115){\scriptsize(a)}
             \put(22,88){\scriptsize(b)}
             \put(22,63){\scriptsize(c)}
             \put(22,37){\scriptsize(d)}
              \put(151,114){\scriptsize(e)}
 \end{picture}
    \caption{The extracted efficiencies and NEPs of the detector at different temperatures with respect to the raw dark count rate. We employ the blocking times of approximately (a) \unit{5}{\micro\second}  at \unit{-60}{\celsius}, (b) \unit{7.5}{\micro\second} at \unit{-70}{\celsius}, (c) \unit{10}{\micro\second} at  \unit{-80}{\celsius} and (d)  \unit{20}{\micro\second} at \unit{-90}{\celsius}. Although the efficiency does not increase by cooling down the detector, the NEP in (e) decreases and thus the detector's sensitivity increases. Measured values are illustrated with symbols, whereas the solid lines present a guide for the eye.}
    \label{fig:3.4}    
\end{figure}
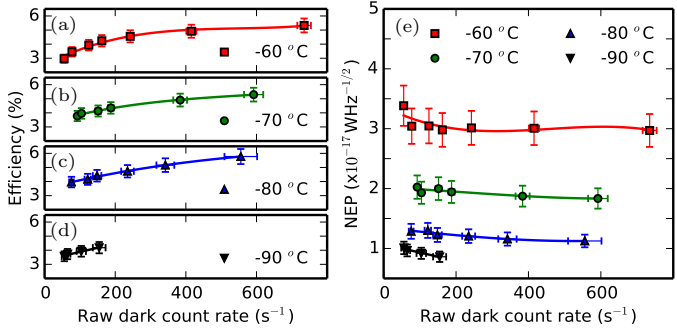

Next, for extracting the detector's efficiency we use a continuous-wave (CW) laser (from Yenista) with a maximal power of \unit{4}{\milli\watt} and tunable between \unit{1520-1600}{\nano\meter} as a light source. The laser beam is attenuated to the few-photon level with a stack of neutral density filters. The weak light is then coupled via a single-mode fiber coupler to the NFADs. We determine the detector efficiency at different raw dark count rates that are fixed by choosing a proper bias voltage. For CW light the detector efficiency can be extracted via \cite{Hadfield2009}
\begin{equation}
\eta = \frac{E_{\textrm{ph}}}{P \mu}\left( \frac{R_{\textrm{det}}}{1-R_{\textrm{det}}\tau} - \frac{D}{1-D\tau} \right),
\end{equation}
in which $E_{ph}$ describes the energy of a photon at \unit{1560}{\nano\meter}, $P$ is the mean power of the CW laser and $\mu=(4.7\pm0.4)\cdot10^{-12}$ the amount of the attenuation. 
In addition to this, $R_{\textrm{det}}$ represents the detected click rate, $D$ is the dark count rate at the selected bias voltage, and $\tau$ the dead time.
We extract the detector's efficiency at different temperatures and at each temperature we adjust the blocking time such that the afterpulsing is suppressed to the same level as with \unit{5}{\micro}{\second} block at \unit{-60}{\degree}. 
 We determine the rates $R_{\textrm{det}}$ and $D$ from the post-selected data and use the blocking time as $\tau$. At each chosen dark count rate we measure the detector's efficiency at different laser powers to ensure that the extracted efficiency remains intact at different photon fluxes and does not vary due to artificial spurious counts. Figs~\ref{fig:3.4}(a-d) illustrate the efficiencies measured at different operating temperatures with respect to the raw dark count rates. We conclude that moderate efficiencies on the order of a few percent can be achieved for CW light.  Nevertheless, the longer blocking times at lower temperatures make the detector useful for applications that require high efficiency but only small count rates, such as triggering on rare or slowly periodic events.

Quite as important as the efficiency is the single photon sensitivity of the detector that can be investigated via the noise equivalent power (NEP) given by \cite{Gisin2002}
\begin{equation}
\textrm{NEP} = \frac{E_{\textrm{ph}}}{\eta}\sqrt{2D}.
\label{eq:NEP}
\end{equation}
The lower the NEP the more sensitive the detector is at the single-photon level. Furthermore, Eq.~(\ref{eq:NEP}) represents the trade-off between the detector dark-count level and the achieved efficiency. Our results  in Fig.~\ref{fig:3.4}(e) show a clear improvement  in NEP when cooling down the detector, nevertheless, remaining rather constant over the range of chosen dark count rates for each temperature. The values obtained here for the NEP follow the tendency of the latest developments reported for avalanche photodetection at telecommunication wavelengths and are gradually approaching those typical in the visible wavelength range at room temperatures although being still one order of magnitude away from what is commercially available for visible light with slight cooling \cite{Buller2010,Warburton2009,Voss2004}.

\begin{figure}[!t]
\includegraphics[width=1.0\columnwidth]{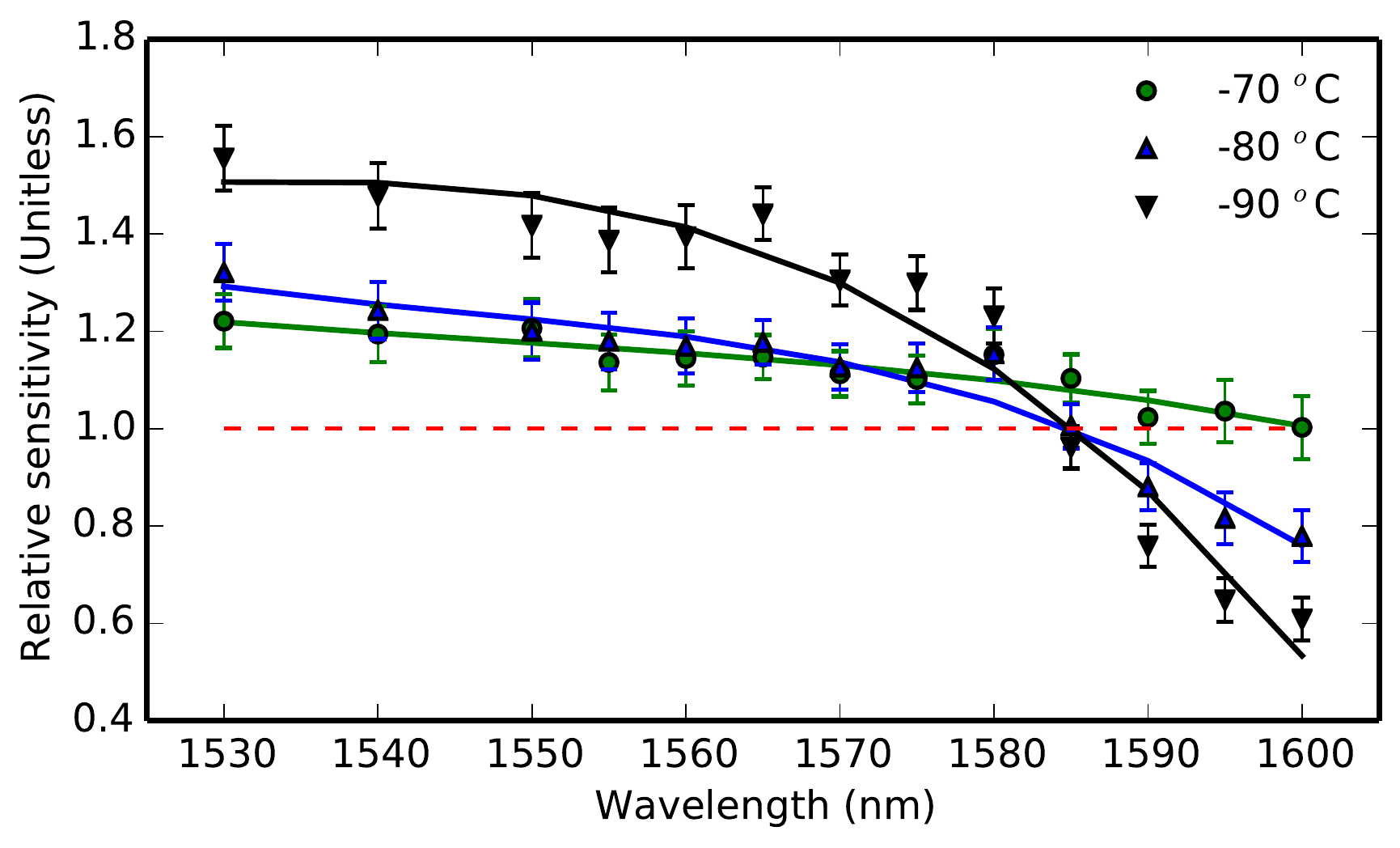}
\caption{The relative sensitivity of the detector with respect to the wavelength at different temperatures. The solid and dashed lines provide guides for the eye.}
\label{fig:3.5}    
\end{figure}

In order to be suitable for the investigation of several different kinds of quantum light sources, a detector with a broad spectral band is highly desired. Since the properties of semiconductor materials are highly dependent on temperature, we investigate the upper cut-off wavelength of our detector that is expected to shift towards shorter wavelengths when the temperature is decreased. 
 For this purpose we measure the relative spectral sensitivity  while trying to keep the raw dark count rates at a constant level between \unit{100-200}{\hertz} at different temperatures. As we are only interested in the relative spectral sensitivity, we choose to employ a fixed blocking time of \unit{5}{\micro\second} at all temperatures, due to which the afterpulsing probability varied from 9.5(8)\%  to 38(6)\% when cooling the detector from \unit{-60}{\celsius} to \unit{-90}{\celsius}. 
We investigate the relative sensitivity at different wavelengths between \unit{1530}{\nano\meter} and \unit{1600}{\nano\meter}. At each wavelength we normalize the  detector efficiency against that at \unit{-60}{\celsius}. From our results in Fig. \ref{fig:3.5} we conclude that in order to gain the best operation cooling down to \unit{-90}{\celsius} is desired only if the light source emits radiation well below \unit{1600}{\nano\meter}. Near this wavelength cooling below \unit{-70}{\celsius} does not improve detectors performance. While we have not directly measured the upper cut-off wavelength, the observed behavior is consistent with its shift towards shorter wavelengths.

Moreover, we test our detectors with a quantum light source that produces photon pairs, typically called signal and idler, via parametric down-conversion (PDC). In our experiment the type-II PDC process is pumped with a picosecond laser (\unit{76}{\mega\hertz} repetition rate, \unit{789\nano\meter} central wavelength, and \unit{0.25}{\nano\meter} bandwidth). 
Our photon-pair source, a \unit{30}{\milli\meter} long pe\-ri\-od\-i\-cal\-ly-poled lithium niobate waveguide, is kept in an oven at a temperature of \unit{186}{\celsius} to reach the desired wavelength region. For coupling  the light  aspheric lenses are employed and after passing through the waveguide the pump beam having CW-equivalent power of approximately \unit{10}{\micro\watt} is separated from signal and idler with a dichroic mirror and a long pass filter. Another dichroic filter is used to block background radiation below wavelengths of \unit{1500}{\nano\meter}. Thereafter, signal and idler are separated in a polarizing beam splitter, coupled to single mode fibers with fixed fiber collimators, and send to the NFADs kept at \unit{-80}{\celsius}. The electric signal of NFADs is discriminated and registered with the TDC. In order to measure the arrival times of signal and idler in a time-resolved manner, the laser clock divided by a factor of 128 is used for synchronization.

\begin{figure}[!t]
\begin{picture}(170,170)%
  \put(0,0){ \includegraphics[width=1.0\columnwidth]{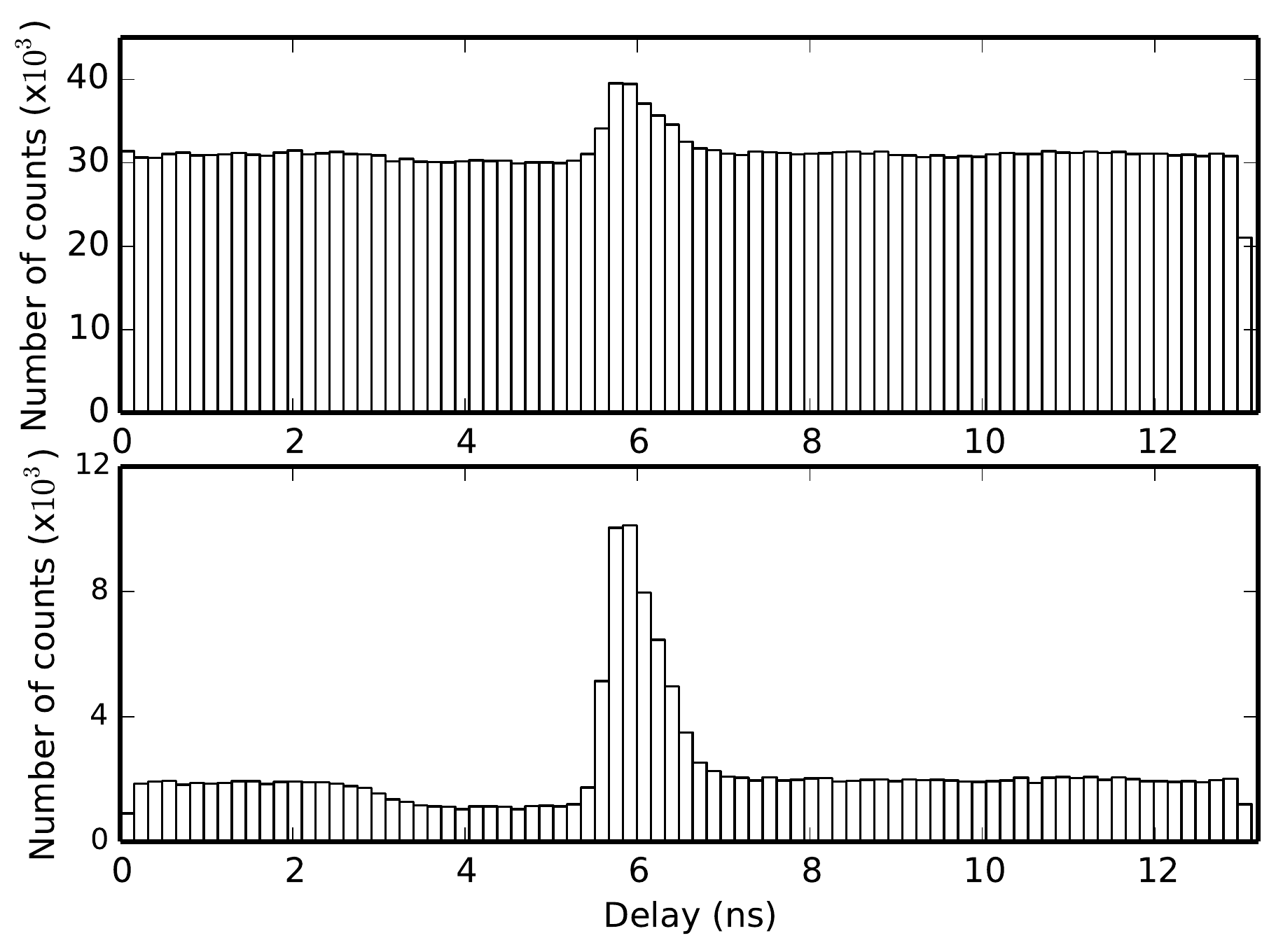}}
   \put(35,161){\scriptsize(a)}
      \put(35,73){\scriptsize(b)}
   \end{picture}
\caption{(a) Histogram of raw single counts in a  PDC marginal in terms of the time delay to the laser clock detected at \unit{-80}{\celsius} with an integration time of \unit{60}{\second}. (b) Histogram of the registered counts, when the data in (a) is post processed with blocking time of \unit{5}{\micro\second} that efficiently removes the large background of spurious counts.  A tight timegating on the order of a nanosecond can further suppress the residual spurious counts.}
\label{fig:3.6}      
\end{figure}

The NFADs are driven with rather low bias voltages such that the raw dark count rates are approximately \unit{310(20)}{\hertz} and \unit{380(30)}{\hertz} for the two detectors. Our results for the raw single counts are shown in Fig.~\ref{fig:3.6}(a) in time resolved manner, which indicates that a large background is detected. The  raw single-count rates measured at the two detectors are \unit{31.5(2)}{\kilo\hertz} and \unit{41.9(4)}{\kilo\hertz}.
We then post-selected the recorded data by employing a blocking time of approximately \unit{5}{\micro\second}. As seen in Fig.~\ref{fig:3.6}(b) this drastically reduces the background of afterpulses and other spurious counts. Further, by applying  narrow time gates of \unit{1.3}{\nano\second}, the post-selected time-gated single-count rates of \unit{810(4)}{\hertz} and \unit{832(4)}{\hertz} are extracted for the two detectors.  This way, the rest of the remaining background can be efficiently suppressed. We wait for afterpulses until a time given by 10\% of the inverse of the post-selected time-gated single-count rate has passed that is approximately \unit{120}{\micro\second} and estimate at the both detectors the afterpulsing probabilities of $0.122(3)$ and $0.120(3)$.

\begin{figure}[!b]
\includegraphics[width=1.\columnwidth]{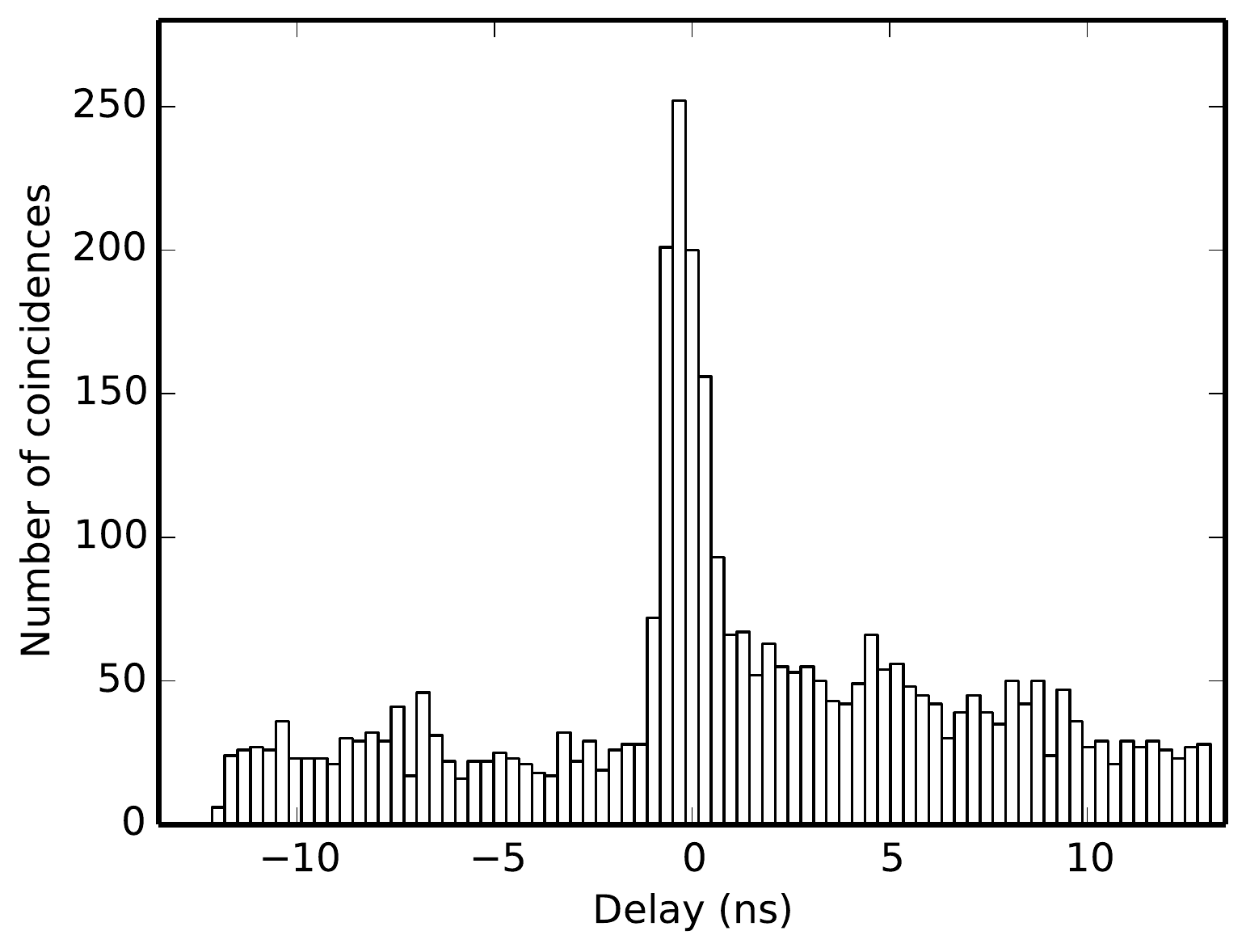}
\caption{Histogram of the raw coincidence counts detected between the two NFADs in terms of the time delay between signal and idler detection times. The time-resolved measurement of coincidences shows a rather low level of background counts that can be easily filtered by applying a tight time gate.}
\label{fig:3.7}      
\end{figure}

Considering coincidence counting, we first investigate the raw coincidence rate in time-resolved manner as shown in Fig.~\ref{fig:3.7}, from which we infer the rate of \unit{58(1)}{\hertz}. By applying a tight time gate, the coincidence count rate drops to \unit{13.3(5)}{\hertz}, and finally by applying blocking it reduces to \unit{12.3(5)}{\hertz}, which seem to be, on the other hand, rather robust against the afterpulsing effects.
Finally, we apply the NFADs in a true state characterization task. We extract the Klyshko efficiency of our PDC process by comparing the number of coincidence counts to single counts, after the reduction of accidental counts from the observed coincidences. This delivers for signal and idler the values $1.48(6)\%$ and $1.52(6)\%$, in which we include only statistical errors.  The coincidence to accidentals ratio of our source is  $1390\pm60$, where the amount of accidentals is estimated via the measured single-count rates.  The state preparation with a fast pump in combination with the free-running detection allows us to drive the source in the extremely low gain regime that is necessary for applications relying on true photon-pair sources.

\section{Conclusions}
\label{sec:4}

Free-running single-photon sensitive detectors for the telecommunication band are of great importance for many quantum optics applications. Still today, most of these are plagued with low detection efficiencies or require special conditions to work. We implemented a highly practical and flexible cooling with liquid nitrogen for two fiber-coupled, commercially available NFADs. With our temperature control extremely stable operation of the detector is possible. We investigated several figures of merit of our detector.  In order to work at its best performance,  we investigated the dark count level and afterpulsing probability. Due to the large afterpulsing we introduced a blocking time that enables us still to measure with a relative high single-photon sensitivity and a moderate detection efficiency. Further, we investigated the spectral boundaries of our detector at the telecommunication wavelengths.  Our results deliver a set of working parameters for the most suitable operation of such a detector in a quantum optics application. Finally, we showed that by applying a blocking time our detector is indeed applicable for time-resolved photon-counting tasks of pulsed light. Despite the drawbacks, our detector shows a great potential, and it can be useful for example in coincidence counting, or even in more sophisticated photon counting applications via time-multiplexing in a fiber network.

\section{Acknowledgements}
\label{sec:5}
We thank Henning Weier (qutools) the support with discriminator boards required for operating our NFADs.  Additionally, we thank Armin Sailer and Gerhard Hendl for helping with the mechanical and electrical construction of the detector and Raimund Ricken for the assistance with the waveguide fabrication. This work was supported in part by the European Research Council (ERC) through project EnSeNa (257531) and the Austrian Science Fund (FWF) through project no. I-2065-N27.


\end{document}